\documentclass[prb, twocolumn]{revtex4-1}
\usepackage{amsmath,amssymb,bm}
\usepackage{graphicx}
\usepackage{epstopdf}
\usepackage{latexsym}
\usepackage{subfigure}
\usepackage{color}
\usepackage{natbib}
      
\begin{document}

\title{Spin nematic ground state of the triangular lattice $S=1$ biquadratic model}
\author{Ribhu K. Kaul}
\affiliation{Department of Physics \& Astronomy, University of Kentucky, Lexington, KY-40506-0055}
\begin{abstract}
Motivated by the spate of recent experimental and theoretical interest in
Mott insulating $S=1$ triangular lattice magnets, we consider a model $S=1$ Hamiltonian on a triangular lattice
interacting with rotationally symmetric biquadratic interactions. We show
that the partition function of this model can be expressed in terms of
configurations of three colors of tightly-packed, closed loops
with {\em non-negative} weights, which allows for efficient quantum
Monte Carlo sampling on large lattices. We find the ground state has spin nematic order, {\em
  i.e.} it spontaneously breaks spin rotation symmetry but preserves
time reversal symmetry. We
present accurate results for the parameters of the low energy field theory, as well as
finite-temperature thermodynamic functions.
\end{abstract}
\maketitle

\section{Introduction}

Magnetism in Mott insulators has blossomed into an exciting
frontier of quantum condensed matter physics on both the
experimental and the theoretical front~\cite{balents2010:spliq}. The
magnetism in such materials is usually modeled by lattice spin
Hamiltonians. These are the simplest many-body problems but yet they
can be notoriously complex. In most cases, one must resort to
uncontrolled approximations to capture the rich landscape of
emergent phenomena.
Given the central role they play, it is of great importance to develop a collection of spin
models that can be studied in the thermodynamic limit in an unbiased
manner. These controlled results can serve as a benchmark for
approximate methods and are sometimes necessary to study certain
non-perturbative phenomena. 
While in one dimension, recent advances in numerical methods have allowed the study of
almost any Hamiltonian~\cite{scholl2005:dmrg}, in two and
higher dimension quantum Monte Carlo  of a small set of ``sign-problem'' free models
provide us with the only unbiased view of the quantum physics of spin
models in the thermodynamic limit~\cite{kaul2013:qmc}.

\begin{figure}[b]
\includegraphics[width=3.3in,trim=0 0 0 280,clip=true]{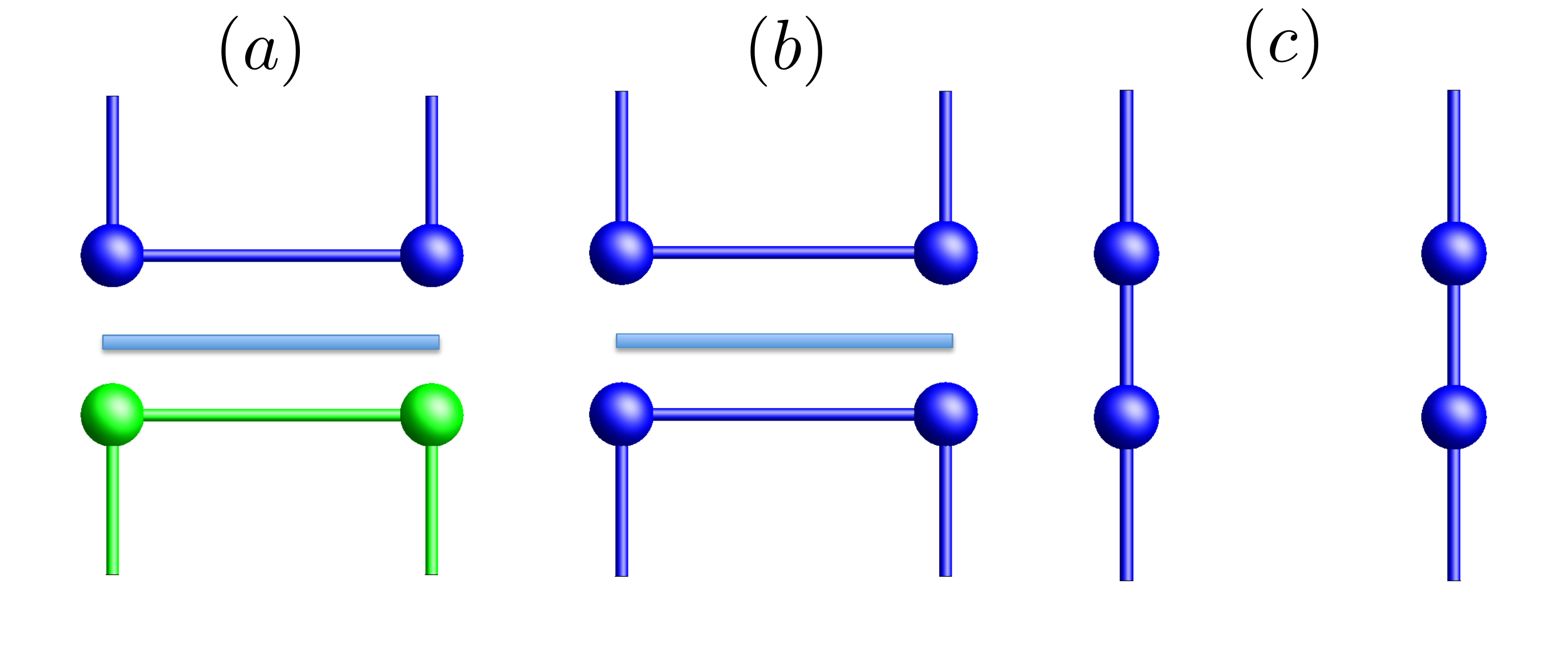}
\caption{\label{fig:loop} Loop configurations. The cartoons show how
  the color ($x,y,z$) of a site and its nearest neighbor evolve in a
  step of imaginary time and how this affects loop decomposition with (a)
  off-diagonal operators, {\em e.g.} $|xx\rangle\langle yy |$ (b)
  diagonal operators, {\em e.g.} $|xx\rangle\langle xx |$, and (c) no
  operator. In the MC updates, updating the colors of the loops converts between (a) and (b). Dividing or joining loops of the same color
converts between (b) and (c). In combination these updates are
ergodic. The horizontal bars denote the action of one of the terms in
the bond operator, Eq.~(\ref{eq:hamterms}).}
\end{figure}

The motivation for the ``sign-problem'' free model we consider here
comes from two recently studied materials with Ni$^{2+}$ ions living on a
triangular lattice, 
NiGa$_2$S$_4$~\cite{nakatsuji2005:nigas} and
Ba$_3$NiSb$_2$O$_9$~\cite{cheng2011:banisbo}. 
Both materials are Mott insulators with the Ni$^{2+}$ ions in a $S=1$ state. In surprising discoveries, it was found that both
materials have gapless ground states but do not realize the 120$^\circ$ magnetically order state
predicted by a semi-classical analysis of the antiferromagnetic
Heisenberg model. Despite further investigations, the low temperature phase in these materials is still
under debate~\cite{nakatsuji2007:nigas}.
Prompted by the experimental work, there has been a number of
theoretical studies of $S=1$ models on the triangular
lattice, using various approximate methods and exact diagonalization
on lattice with up to 21 sites. 
Depending on the model or method chosen, researchers have found magnetic states,
proximity to quantum criticality~\cite{chen2012:s1tri}
various spin nematics~\cite{laeuchli2006:nigas,tsunetsugu2006:nigas,bhattacharjee2006:nigas} and spin liquids~\cite{grover2011:s1,xu2012:s1tri}.
One popular theoretical rationalization for the unusual
experimental behavior is the presence of a strong biquadratic
interaction~\cite{stoudenmire2009:nigas}.

Inspired by the extensive experimental and theoretical work we study a $S=1$
model on the triangular lattice with
pure biquadratic exchange. Remarkably, we find that this Hamiltonian does
not suffer from the sign problem, and exploit this
fact to study systems without any approximation on lattices with more than $10^4$ spins.  Using our numerical
results we confirm that the biquadratic model has a {\em spin nematic}
ground state,  consistent with previous approximate
studies and small cluster exact diagonalization studies of the same Hamiltonian~\cite{bhattacharjee2006:nigas,laeuchli2006:nigas}.
We then make accurate estimates for  the low energy parameters of this model, thanks to the large system sizes
made accesible by our method. 

\section{ Model}
 
The model we are interested in can be expressed in terms of the spin-1 Pauli matrices $\vec S$ as,
\begin{equation}
\label{eq:biq}
\hat H = - K\sum_{\langle ij \rangle} \left (\vec S_i \cdot \vec S_j\right ) ^2
\end{equation}
where the sum is taken over pairs of nearest neighbor sites of a
triangular lattice. 
It is well known that the pure biquadratic model
has a staggered SU(3) symmetry on bipartitie lattices. On the
non-bipartite triangular lattice studied here it has only the symmetry of
spin rotations. 
In the usual labeling of the eigenstates on a site, the states $|1\rangle, |0\rangle$ and $|\bar 1 \rangle$ are eigenstates of $S_z$ with eigenvalues $1,0$ and $-1$. 
When written in the form above, the model, Eq.~(\ref{eq:biq}) appears to have off-diagonal matrix elements with both positive and negative signs, violating the Marshall sign rule. 
The mappings to follow are most transparent in the orthonormal basis of eigenstates of $S_x,S_y$ and $S_z$ with eigenvalue $0$. These are,
\begin{eqnarray}
|x\rangle&=& \frac{1}{\sqrt{2}}(|1\rangle -|\bar 1 \rangle)\nonumber\\
|y\rangle&=& \frac{i}{\sqrt{2}}(|1\rangle +|\bar 1 \rangle)\nonumber\\
|z\rangle &=& |0\rangle
\end{eqnarray}
It is possible to verify by inspection that in terms of these basis states the biquadratic operator can be re-written simply as a projector,
\begin{eqnarray}
\label{eq:hamterms}
\left (\vec S_i \cdot \vec S_j\right ) ^2 -1 =3 |\mathcal{S}\rangle
\langle \mathcal{S} |
\end{eqnarray} 
where $|\mathcal{S}\rangle = \frac{|xx\rangle+|yy\rangle+|zz\rangle
}{\sqrt{3}}$. In this basis it is transparent that the Hamiltonian,
Eq.~(\ref{eq:biq}), satisfies the Marshall sign criteria, all
off-diagonal matrix elements are negative or zero, {\em even on
  non-bipartite lattices}. We note in that although as mentioned above the model does
not have all non-positive off diagonal matrix elements in the standard $S_z$ basis,
this situation can be remedied by the phase rotation $| 0
\rangle \rightarrow i |0\rangle$, in which case  $|\mathcal{S}\rangle
= \frac{|00\rangle+|\bar 11\rangle+|1\bar 1\rangle
}{\sqrt{3}}$, then from Eq.~(\ref{eq:hamterms}) it has all its matrix
elements non-positive.

\begin{figure}[t]
\includegraphics[width=3.5in]{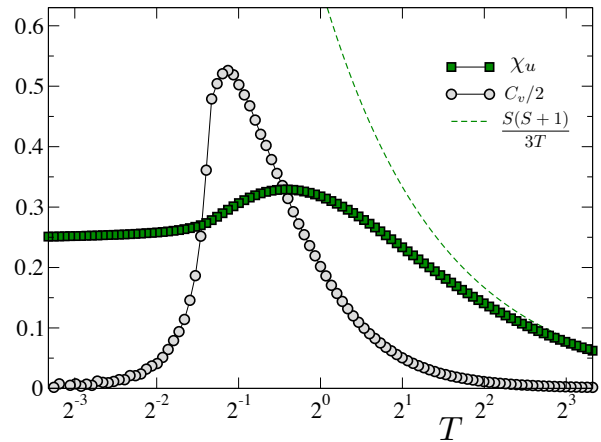}
\caption{\label{fig:thermo} Finite-temperature specific heat, $C_v$
  (circles) and uniform
  susceptibility, $\chi_u$, (squares) of the $S=1$ biquadratic model, Eq.~(\ref{eq:biq}) on a $64\times
  64$ triangular lattice. The results are representative of the
  thermodynamic limit. The single spin Curie law $S(S+1)/3T$ is
  shown for comparison as a dashed line. We have plotted $C_v/2$ to
  make both data sets fit on the same scale. }
\end{figure}

The Marshall condition  implies that our model Eq.~(\ref{eq:biq})  has a
positive definite path integral, if  we use the $|x\rangle, |y \rangle,
|z\rangle$ basis. For concreteness we use the stochastic series
expansion (SSE) method to change from Hamiltonian to path integral at
finite temperature $T=1/\beta$, in
which ${\mathcal Z}=\sum_n \frac{(-\beta)^n}{n!}{\rm Tr}(\hat H
^n)$~\cite{sandvik2010:vietri} (The mapping below goes through equally
well in the usual Trotter path integral approach). An SSE
configuration is specified by assigning to each $\hat H$ in the trace
one of the terms of Eq.~(\ref{eq:hamterms})  that contributes to the Hamiltonian and its location on the lattice, {\em e.g.}
$|xx\rangle_{ij} \langle zz|_{ij}$ (see Fig.~\ref{fig:loop}). The partition function is a sum
over all SSE configurations, which at any given step in the evaluation
of the trace is described by specifying the basis state the system is
in at that time slice, {\em i.e.}, by stating which one of the three colors,
$x,y$ or $z$, is on each
lattice point. Colors can change through the action of the
off-diagonal terms in Eq.~(\ref{eq:hamterms}) in which two neighboring
lattice points with the same color can both switch at the next time slice
to a new color [see Fig.~\ref{fig:loop}(a)]. It can now be shown that
every SSE configuration is identified with a unique loop decomposition: loops are constructed
by starting at a point in space-time and moving along the world-line in
the time direction until a bond operator is encountered at which point one
traverses to the neighboring site on the bond and switches the
direction of motion until the loop closes; loops join space-time
points with the same color and every space-time point is connected to
only one loop  (see Ref.~\cite{syljuasen2002:dirloop} for related details). Putting the entire imaginary time history together, it is easy to see that every
world-line configuration maps uniquely to a configuration of closed
tightly packed loops, each of which is assigned one of three colors,
$x,y$ or $z$. The weight of any loop configuration from the stochastic
series expansion is then given by
the simple formula,
\begin{equation}
\label{eq:wght}
{\mathcal Z} = \sum_{ l} \frac{(\beta K)^{n_l}}{n_l!} 3^{{\cal N}_l},
\end{equation}
where the sum over $l$ is a sum over all closely packed loop
configurations, $n_l$ is the number of operator insertions (shown in
Fig.~\ref{fig:loop} as horizontal blue bars) and
$3^{{\cal N}_l}$ is an entropic factor that counts the number of ways
colors can be assigned to the ${\cal N}_l$ loops in $l$.

\section{Quantum Monte Carlo} 

We implement the described mapping of
Eq.~(\ref{eq:biq}) to a loop model
 and sampled the SSE configurations with Monte Carlo updates. Loop configurations
can be updated by joining loops of the same color, dividing a loop
into two smaller loops and changing the color of a loop (see the
caption of Fig.~\ref{fig:loop}). Such updates
are ergodic and can be chosen with the proper weights to satisfy
detailed balance. We note here that past quantum Monte-Carlo work has looked at the
biquadratic interaction using a different 
algorithm in which $S=1$ operators are decomposed into two $S=1/2$
operators as well as a loop algorithm similiar to the one described
here, but in the $S^z$ basis. These studies were however restricted to bipartite lattices (1-d chains, 2-d
square and 3-d cubic lattices)~\cite{kawashima2004:review,harada2001:blbq,harada2002:blbq}.

One can now derive expressions in the loop language for various
quantities of interest in the spin model. The total energy $E$ and the
specific heat $C_v\equiv \frac{1}{N_s}\frac{dE}{dT}$ can be
measured in the usual manner of the SSE~\cite{sandvik2010:vietri}. The uniform susceptibility
$\chi_u \equiv \frac{\beta}{N_s}\left \langle \left (\sum_{i=1}^{N_s} S^z_i\right
  )^2 \right\rangle$ can be shown to be exactly $\frac{2}{3}\left \langle
\sum_i \left (W^\tau_i\right )^2\right \rangle $ where $W^\tau_i$ is
the temporal winding number and $i$ is the index which sums over the
loops identified with a given configuration in space-time. The spin
stiffness is related to the well known winding number estimator
$\rho_s= T \left \langle
\sum_i \left (W^x_i\right )^2\right \rangle$, where $W^x$ is the
spatial winding number.

\begin{figure}[t]
\includegraphics[width=3.5in,trim=20 0 0 280,clip=true]{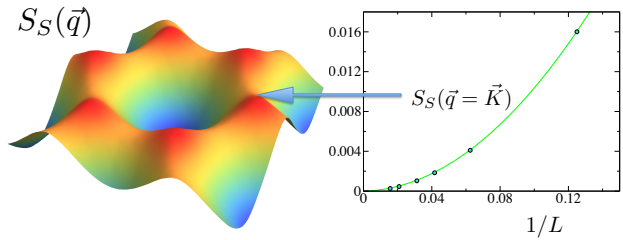}
\caption{\label{fig:struct} Equal time spin structure factor.
  (left) Re$[S_S(\vec q)]$ shown in the full first Brillouin zone of the
  triangular lattice on a $32\times 32$ system at $\beta=32$. (right)
  the peak height of the structure factor at the corner of the BZ (so-called ``K''-point)  as
  a function of inverse system size. The solid circles are
  the QMC data and the solid line is a numerical fit that shows the
  spin order vanishes in the thermodynamic limit.}
\end{figure}

We begin by studying the specific heat and the
susceptibility of our magnet as it is cooled to its ground state. The
data in Fig.~\ref{fig:thermo} on a $64\times 64$ system is representative of the thermodynamic limit. 
The peak in $C_v(T)$ at $T\approx 0.5$ signals the lifting of the
extensive entropy of the spins in the high-temperature phase. The observed simultaneous
saturation of $\chi_u(T)$ is the classic behavior expected of an
anti-ferromagnet. Even so, a study of the spin structure factor
$S_S(\vec q) \equiv \frac{1}{3L^2} \sum_r e^{ik \cdot r}\langle \vec S(r) \cdot
\vec S(0)\rangle$, shows no peaks that scale to a finite value in the
thermodynamic limit, indicating the absence of simple
magnetic long range order as shown in Fig.~\ref{fig:struct}.

\begin{figure}[t]
\includegraphics[width=3.7in,trim=70 0 0 0,clip=true]{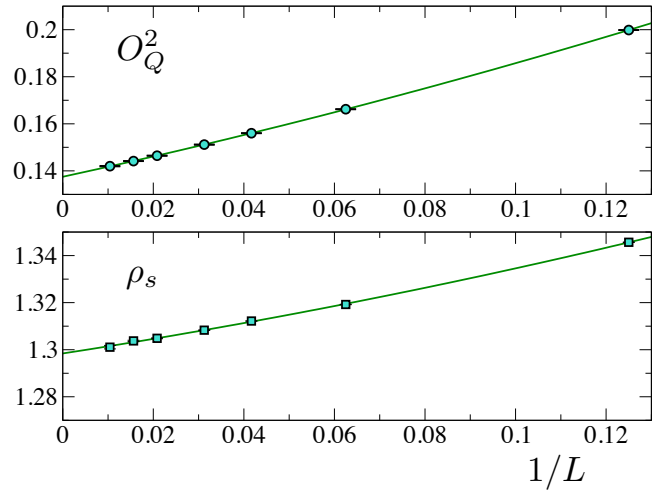}
\caption{\label{fig:fss} Finite-size scaling at fixed $K\beta/L=1$.  The
  upper panel shows finite size scaling of the data for the nematic
  order parameter, which in the thermodynamic limit extrapolates to
  $O_Q^2=0.1376(2)$. The lower panel shows the extrapolation of the
  spin stiffness to a value $\rho_s = 1.298(2)$. }
\end{figure}

A natural guess for a magnetic state with no Bragg peaks in the spin
structure factor is a {\em spin nematic}. Such a state breaks spin
rotational symmetry but preserves time reversal symmetry. The order
parameter for the spin nematic is a traceless symmetric matrix,
\begin{equation}
\hat Q_{\alpha\beta} = \frac{\hat S_\alpha \hat S_\beta +\hat S_\beta \hat S_\alpha}{2} -
\frac{2\delta_{\alpha\beta} }{3}.
\end{equation}
where $\hat S_\alpha$ are components of the $S=1$ operator. Since the
order parameter is even in the spin operators, its condensation does
not imply breaking of time reversal. In order to test for this order we measure
the correlation function $C_{Q}(r) =\sum_{\alpha} \langle
\hat Q_{\alpha\alpha} (r)\hat Q_{\alpha\alpha}(0)\rangle$, which is
easy to measure in our basis since it is diagonal. It is possible to
show that the O(3) invariant correlation function is simply related by
a constant, {\em i.e.}, $C_Q(r)=\frac{2}{5} \sum_{\alpha\beta} \langle
\hat Q_{\alpha\beta} (r)\hat Q_{\beta\alpha}(0)\rangle$.
Fig.~\ref{fig:fss}(a) shows a plot of $O_Q^2= \frac{1}{L^2}\sum_r C_Q(r)$ versus
$1/L$, keeping the ratio of $K\beta/L=1$ fixed so that we probe the
ground state behavior. The finite
value for $O_Q^2$ in the $T=0$ and thermodynamic limit (combined with no condensation in the
magnetic structure factor) proves unambiguously that the ground state of our model,
Eq.~(\ref{eq:biq}), is a {\em spin nematic}. It is interesting to ask
how much our ground state deviates from a simple product nematic
state with no quantum fluctuations, a wavefunction for which is simply, $|\Psi_{\rm prod}\rangle = \prod_i
|S^z=0\rangle_i$. In this state the $\langle Q_{\alpha\beta}\rangle=
\delta_{\alpha\beta}/3 - \delta_{\alpha z}\delta_{\beta z}$, from
which we conclude that $(O^{\rm prod}_Q)^2 =4/15\approx
0.26666...$. This is the maximum value the order parameter can
take, if there are no fluctuations. From the extrapolation in Fig.~\ref{fig:fss} we find that
$(O_Q)^2 =0.1376(2)$ in the real ground state, the reduction of about
50\% being due to the effect of quantum fluctuations.

\begin{figure}[t]
\includegraphics[width=3.6in,trim=80 0 0 0,clip=true]{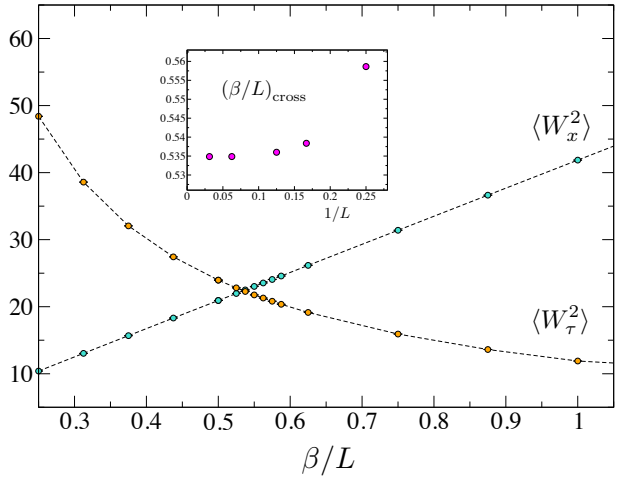}
\caption{\label{fig:vel} Determination of the spin wave velocity, $c$.
In the main panel we show data for the $\langle W_x^2\rangle$
and $\langle W_\tau^2\rangle$ on an $L=32$ system as a function of $\beta/L$. In the inset we
show how the $\beta/L$ value at the crossing of the types of plots in the main panel scales with system
size. These are finite-size estimates for $1/c$, from which we obtain
$c=1.869(4)$ in the thermodynamic limit.}
\end{figure}

The breaking of the continuous O(3) spin rotational symmetry results
in gapless spin waves. Analogous to past work on the N\'eel state~\cite{chakravarty1988:qaf} the
low energy effective theory of the  Goldstone
bosons at leading order is characterized completely by just two parameters, the
spin stiffness, $\rho_s$, and the spin wave velocity, $c$. The
stiffness can be estimated by the usual winding number estimator. Fig.~\ref{fig:fss} (b)
shows a plot of $\rho_s$ as a function of $1/L$, at fixed $K\beta/L=1$. A finite
thermodynamic value for $\rho_s$ provides further evidence for gapless
spinful excitations at long wavelengths.  The spin wave velocity is a number that converts the units of space
into the units of time. An estimator for the spin wave velocity, $c$,
is hence the ratio of
$L/\beta$ at which the anisotropic 2+1 dimensional system behaves cubic. We compute
this ratio for each $L$ by adjusting $\beta$ until $\langle
{W_x}^2 \rangle = \langle {W_\tau}^2 \rangle$. Then we carry out a
thermodynamic extrapolation of the finite size $L/\beta$ data to
obtain an estimate for $c$. The data for the
winding numbers for a $32\times 32$ system is shown as an example in
main panel of Fig.~\ref{fig:vel}.
We fit polynomials through this data and estimate the location of the
crossing point for each $L$ studied. The value of $L/K\beta$ at the crossing is then plotted
versus $1/L$ in the inset. Extrapolating to the thermodynamic
limit we obtain that the spin wave velocity, $c=1.869(4)$. The
estimates of the order parameter $Q_{\alpha\beta}$, the spin stiffness
$\rho_s$ and the spin wave velocity $c$, give a complete description
of the low energy effective field theory to leading order.

\section{Summary}

In conclusion, we have provided a model $S=1$ hamiltonian with
biquadratic interactions that is
sign-problem free on non-bipartite lattices. We have shown how the
partition function of this model maps to a remarkably simple loop model with
positive weights, see Eq.~(\ref{eq:wght}). In this publications we studied how this mapping has allowed for a
detailed study of the spin nematic ground state on the triangular
lattice relevant to many recent experimental and theoretical
works and provided high precision estimates for the ground state paramaters. Extensions of the current work to carry out unbiased numerical
studies of the role of disorder in quantum spin nematics
and deconfined critical points out of the nematic phase~\cite{harada2007:deconf,grover2007:deconf} are exciting directions
for future research. Another field of experiments in which the nematic
phase for $S=1$ spins has been discussed is in ultra-cold atoms in
optical lattices; it is hoped that our high precision results may
be useful for guiding and testing experimental efforts in this context~\cite{imambekov2003:s1}.

The author is grateful to Matt Block and Michael Levin for very valuable
input, and to Andreas L\"auchli for providing ED data used to
test the QMC code. The work was supported in part by NSF DMR-1056536.

\bibliography{/Users/rkk/OPPIE/Physics/PAPERS/BIB/career.bib}

\end{document}